\documentclass[prl,longbibliography,twocolumn,nopacs,preprintnumbers,notitlepage,amsmath,amssymb,superscriptaddress]{revtex4-2}

\usepackage{amsmath}
\usepackage{amssymb}
\usepackage{amsbsy}
\usepackage{graphicx}
\usepackage{color}

\usepackage[normalem]{ulem}

\usepackage{enumerate}
\usepackage{mathtools}

\newcommand{\moy}[1]{\left\langle #1 \right\rangle}

\newcommand{\XX}[0]{\boldsymbol{X}}

\newcommand{\gt}[0]{\widetilde{g}}
\newcommand{\dd}[0]{\mathrm{d}}

\newcommand{\zz}[0]{\mathbf{0}}

\newcommand{\ee}[0]{\boldsymbol{e}}
\newcommand{\rr}[0]{\boldsymbol{r}}

\newcommand{\RR}[0]{\boldsymbol{R}}

\definecolor{darkblue}{rgb}{0,0,0.6}
\definecolor{darkred}{rgb}{0.6,0,0}
\usepackage[colorlinks=true,urlcolor=darkblue,citecolor=darkblue,linkcolor=darkred]{hyperref}

\begin{document}

\title{Absolute negative mobility of an active tracer in a crowded environment}

\author{Pierre Rizkallah}
\affiliation{Sorbonne Universit\'e, CNRS, Laboratoire de Physico-Chimie des \'Electrolytes et Nanosyst\`emes Interfaciaux (PHENIX), 4 Place Jussieu, 75005 Paris, France}

\author{Alessandro Sarracino}
\affiliation{Dipartimento di Ingegneria, Universit\`a della Campania Luigi Vanvitelli, 81031 Aversa (CE), Italy}

\author{Olivier B\'enichou}
\affiliation{Sorbonne Universit\'e, CNRS, Laboratoire de Physique Th\'eorique de la Mati\`ere Condens\'ee (LPTMC), 4 Place Jussieu, 75005 Paris, France}

\author{Pierre Illien}
\affiliation{Sorbonne Universit\'e, CNRS, Laboratoire de Physico-Chimie des \'Electrolytes et Nanosyst\`emes Interfaciaux (PHENIX), 4 Place Jussieu, 75005 Paris, France}

\date{\today}

\begin{abstract}

  Absolute negative mobility (ANM) refers to the situation where the average velocity of a driven tracer is opposite to the direction of the driving force. This effect was evidenced in different models of nonequilibrium transport in complex environments, whose description remains effective.  Here, we provide a microscopic theory for this phenomenon. We show that it emerges in the model of an active tracer particle submitted to an external force, and evolving on a discrete lattice populated with mobile passive crowders. Resorting to a decoupling approximation, we compute analytically the velocity of the tracer particle as a function of the different parameters of the system, and confront our results to numerical simulations.  We determine the range of parameters where ANM can be observed, characterize the response of the environment to the displacement of the tracer, and clarify the mechanism underlying ANM and its relationship with negative differential mobility (another hallmark of driven systems far from the linear response).

\end{abstract}

\maketitle

\emph{Introduction.---} Predicting the response of a tracer particle to an external driving is a central challenge in statistical physics \cite{Kubo1966,Marconi2008}. In particular, within the framework of transport phenomena, the object of study is the relation between the applied external field and the induced displacement. This so-called `force-velocity relation' can  display a number of striking anomalies, typically when the tracer has a nonequilibrium dynamics and evolves in a complex environment. One of the most intriguing behaviors is the onset of an inverse current, which is opposite to the driving force. In  the specific context of particle transport, this effect is known as absolute negative mobility (ANM).  
This effect was originally observed in experiments with colloidal particles in a
microfluidic system under the action of an
alternating-current electric field, and has been mostly applied for particle sorting~\cite{ros2005absolute,eichhorn2010negative,slapik2019tunable}. Negative response phenomena were also observed in other fields for other observables: in the context of strongly correlated electron systems,  analogous behaviors have been exploited for directly  measuring the viscosity-to-resistivity ratio~\cite{levitov2016electron}, and in Josephson junctions~\cite{nagel2008observation}.

From a theoretical perspective, ANM has been evidenced in models of Brownian particles subjected to a periodic time-dependent force~\cite{kostur2006forcing,machura2007absolute,speer2007transient},
or to dichotomous noise in an asymmetric spatial structure~\cite{eichhorn2002brownian}.
Other examples of systems exhibiting ANM are a random walk with persistence and correlated step-size~\cite{cleuren2002random}, a driven tracer on a one-dimensional lattice with hard-core and exchange interactions~\cite{cividini2018driven},  Hamiltonian coupled transport \cite{wang2020inverse}, or an inertial probe in two dimensions advected by a steady velocity field~\cite{sarracino2016nonlinear, cecconi2017anomalous, ai2018giant}. More recently, ANM has been investigated in the context of active matter, within the Active Ornstein-Uhlenbeck model~\cite{cecconi2018anomalous}, or for active polymers~\cite{wu2022absolute} and Janus particles in convective flows~\cite{li2021enhanced}.

\begin{figure}[b]
    \centering
    \includegraphics[width=\columnwidth]{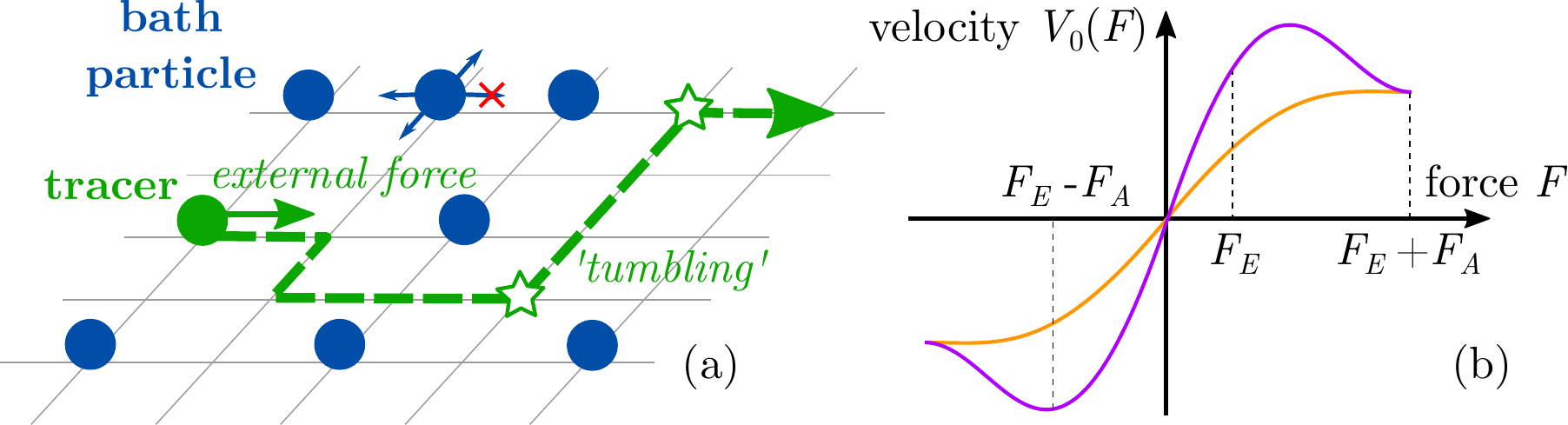}
    \caption{(a) Sketch of the system under study: a tracer particle (in green), which is active and submitted to an external driving, evolves in a bath of passive crowders. (b) Typical force-velocity curve of a passive tracer with and without negative differential mobility (in purple and orange, respectively).}
    \label{fig:model}
\end{figure}

However, in the different analytical models that were designed to study the conditions under which ANM may occur, the environment of the tracer is  described implicitly (through effective interaction potentials or through a background flow, for instance). Here, we fill this gap and propose a minimal microscopic model in which ANM occurs. Our model relies on the symmetric exclusion process: we consider  an active tracer particle submitted to an external force, which evolves in a dynamical environment of mobile passive crowders on a lattice, whose dynamics is accounted for explicitly.  This model thus belongs to the important class of exclusion processes, which are paradigmatic models of nonequilibrium statistical physics, and which received considerable attention, both in 1D \cite{Chou2011a,Mallick2015}, and in higher dimensions \cite{Nakazato1980, Kehr1981, Tahir-Kheli1983a, Leitmann2013, Leitmann2017, Cividini2017a, Pigeon2017, Benichou2018}.
On top of providing an explicit description of the environment of the tracer, which allows us to characterize the response of the environment to the displacement of the tracer, this model is analytically tractable, gives accurate results in a wide range of parameters, and elucidates the conditions under which ANM is observed.
We also present a qualitative argument valid at low density, which explains the main phenomenon in terms of the relevant characteristic timescales. In particular, we show how ANM emerges from the trapping of the tracer particle by the passive crowders. Finally, our approach clarifies  the relationship between ANM and negative differential mobility (another hallmark of driven systems far from the linear response).

\emph{Model.---} We consider a $d$-dimensional cubic lattice ($d\geq 2$) of unit spacing, with base vectors $\ee_1,\dots,\ee_d$ (we use the convention $\ee_{-\mu} = -\ee_{\mu}$). The bath particles perform continuous-time symmetric random walks: at times drawn randomly from an exponential clock of average $\tau^*$, they pick one of their $2d$ neighboring sites at random, and jump onto if it is free (otherwise, the jump is not performed). The tracer particle also performs jumps onto neighboring sites, but is under the influence of two forces: (i) a constant external force, pointing in direction $\ee_1$, and of intensity $F_E$, (ii) an active force of intensity $F_A$ and whose direction $\ee_\chi$ ($\chi \in \{\pm 1,\dots,\pm d\}$) changes randomly at exponentially distributed times of average $\tau_\alpha$. This active force represents the `propulsion' of the particle (for instance that of a microswimmer, such as a bacteria or an active colloid), whereas the force $F_E$ represents some external driving imposed on the active particle, that may originate from a solvent flow or a magnetic field, for instance (see Fig. \ref{fig:model}(a) for a sketch of the model). The tracer jumps in direction $\ee_{\mu}$ if the target site is empty, with rate ${p_\mu^{(\chi)}}/{\tau}$, where  $ p_\mu^{(\chi)} = \frac{\exp \left[ (F_A\ee_{\chi} + F_E\ee_1) \cdot \ee_\mu /2 \right]}{Z}$. The normalisation factor $Z$ is such that $\sum_\mu p_\mu^{(\chi)} = 1$ (we use the shorthand notation $\sum_\mu \equiv \sum_{\mu \in\lbrace \pm 1, ..., \pm d\rbrace }$). Finally, the hardcore interactions between the particles are enforced by the condition that there can only be one particle per site. The position of the tracer at time $t$ is denoted by $\XX_t$, and we will be interested in its projection along the direction of the external force $X_t \equiv \XX_t \cdot \ee_1$. The velocity  reached by the tracer (along direction $1$) in the stationary state will be denoted by $   V\equiv \lim_{t\to\infty} \frac{\dd \langle X_t \rangle}{\dd t }.$

\begin{figure}
\begin{center}
\includegraphics[width = 0.49\columnwidth]{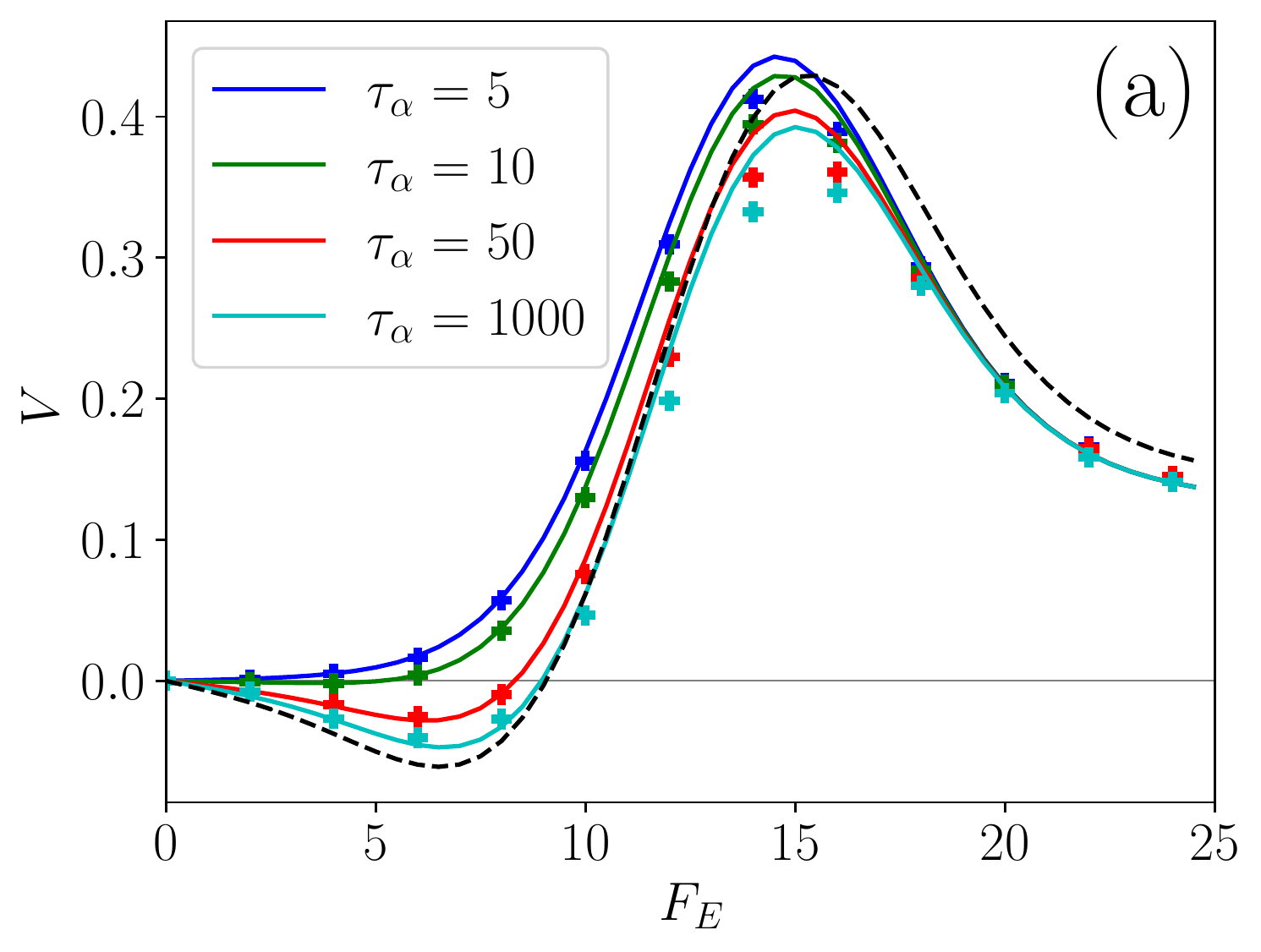}
\includegraphics[width = 0.49\columnwidth]{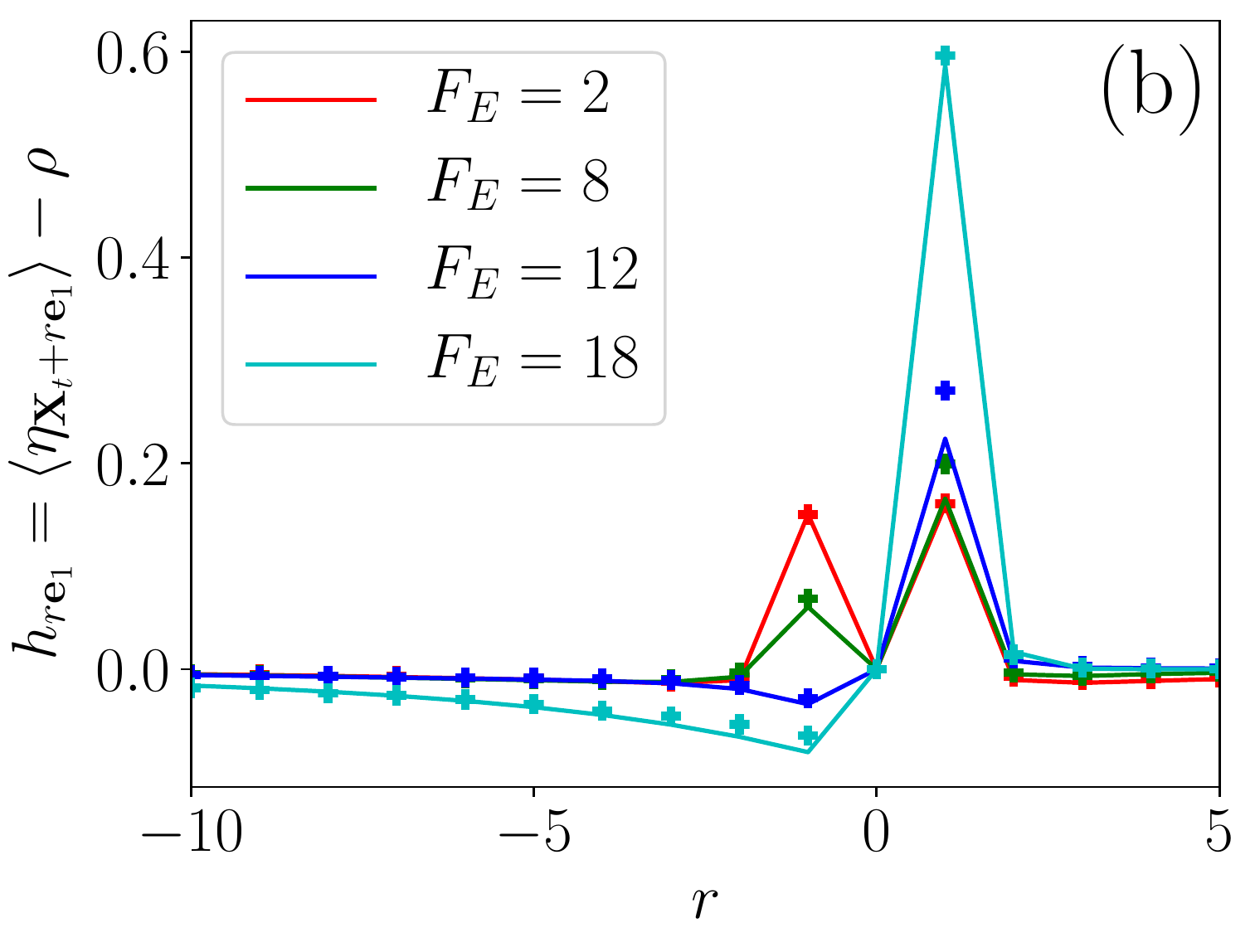}
\caption{ (a) Stationary velocity of the tracer particle along the direction of the external force $F_E$ on a 2D lattice. Lines: decoupling approximation [Eqs. \eqref{eq:speed} and \eqref{eq:eqk_decoupling}]; symbols: Monte Carlo simulations \cite{SM}; dashed line: qualitative argument in the limite of infinite persistence (Eq.~\eqref{eq:qualitativeSpeed} for $\tau_\alpha = \infty$). (b) Density profiles (relatively to the reference value $\rho$) along the direction of $F_E$ in the frame of reference of the tracer, as a function of the distance to the tracer $r$ (here  $\tau_\alpha = 50$). In both plots, the parameters are $\rho=0.1, \tau^* = 30, \tau = 1, F_A=12$. } 
\label{fig:decMC}
\end{center}
\end{figure}

\emph{Simple argument for ANM.---}
\label{simple_arg}
Before going into the details of our analytical derivation, we present a simple argument to explain the emergence of ANM in the model described above, and its relation to negative differential mobility (NDM). The latter refers to the situation where a particle submitted to a constant external force $F_E$ may display a velocity which decreases with the intensity of the force while remaining positive \cite{baerts2013frenetic,Benichou2014,Leitmann2013,baiesi2015role,Benichou2015c, Benichou2018,reichhardt2017negative,zia2002getting,maes2017non}.

For illustration, we consider the simple situation where the tracer is submitted to an active force $F_A$ that may only point in directions $\pm \ee_1$. In the limit where the persistence time is greater than other timescales, the average velocity of the tracer can therefore be estimated as the average of the velocities conditioned on these two states  $V \simeq \frac{1}{2} [V_0(F_E+F_A)+V_0(F_E-F_A)]$, where $V_0(F)$ is the stationary velocity of a passive particle (i.e. with $F_A=0$ and/or $\tau_\alpha=0$) submitted to an external force $F$. We assume that $F_E>0$, and we first consider the case where the tracer does not display negative differential mobility. Its velocity is then a monotonic function of the force undergone by the tracer (see Fig. \ref{fig:model}(b) for a sketchy representation of the force-velocity curves). In this situation, it is clear that ${V}$ will be of the sign of $F_E$, and no absolute negative mobility can be observed. However, when the tracer displays negative differential mobility, one may observe the situation where $|V_0(F_E-F_A)|>|V_0(F_E+F_A)|$, therefore resulting in a situation where average velocity ${V}$ is negative although $F_E>0$. 

These simple considerations clarify the relationship between ANM and NDM. To summarize, the velocity of the active tracer can be understood as an average over the velocities conditioned over the different possible orientations of the active force. If this conditional velocity is a non-monotonic function of the force (NDM), the average velocity can become negative (ANM).

\emph{Results.---}
We now turn to the details of our analytical approach. We define $P_\chi(\RR,\eta;t)$ as the probability that the tracer is at position $\RR$, the active force in direction $\ee_{\chi}$ and the environment in configuration $\eta=\{\eta_{\rr}\}$ (where  $\eta_{\rr}$ is a random variable equal to 1 if there is a bath particle on the site $\rr$ and 0 otherwise) at time $t$. This quantity obeys the following master equation:
\begin{equation}
2d\tau^*\partial_t P_\chi(\RR,\eta;t) = \mathcal{L}_\chi P_\chi - \alpha P_\chi +\frac{\alpha}{2d-1} \sum_{\chi'\neq\chi} P_{\chi'},
\label{eq:master}
\end{equation}
where $\alpha = {2 d \tau^*}/{\tau_\alpha}$ is a dimensionless rate of reorientation of the active force, and  $\mathcal{L}_\chi$ is the evolution operator when the active force is in direction $\ee_{\chi}$:
\begin{eqnarray}
\mathcal{L}_\chi P_\chi=&&
  \sum_{\nu=1}^d\sum_{{\rr}\neq\RR-\ee_\nu,\RR} \left[ P_\chi(\RR,\eta^{{\rr},\nu};t)-P_\chi(\RR,\eta;t)\right]\nonumber\\
&&+\frac{2d\tau^*}{\tau}\sum_{\mu} p_\mu^{(\chi)} 
[\left(1-\eta_{\RR} \right)P_\chi(\RR-\ee_{\mu},\eta;t) \nonumber\\ && ~~~~~~~~~~~~~~~~~-\left(1-\eta_{\RR+\ee_{\mu}}\right)P_\chi(\RR,\eta;t)].
\label{eq:evolution_chi}
\end{eqnarray}
We denote by $\eta^{{\rr},\nu}$ the configuration obtained from $\eta$ by switching the occupations of sites $\rr$ and $\rr + \ee_\nu$. The first term in Eq.~\eqref{eq:evolution_chi} accounts for the jumps performed by bath particles, whereas the second term accounts for the jumps performed by the tracer.

From Eq. \eqref{eq:master}, we derive the expression of the average velocity (by multiplying Eq.~\eqref{eq:master} by $X_t$ and averaging over all positions $\RR$ and lattice configurations $\eta$):
\begin{equation}
\dfrac{\dd \moy{X_t}}{\dd t} = \dfrac{1}{2d \tau}\sum_\chi \left\lbrace p_1^{(\chi)} \left[1-k_{\ee_1}^{(\chi)}\right]-p_{-1} ^{(\chi)}\left[1-k_{\ee_{-1}}^{(\chi)}\right]  \right\rbrace ,
\label{eq:speed}
\end{equation}
where $k_{\rr}^{(\chi)} = \moy{\eta_{\XX_t+\rr}}_\chi = 2d \sum_{\RR,\eta} \eta_{\RR + \rr} P_\chi(\RR,\eta;t)$ is the average occupation of position $\rr$ in the frame of reference of the tracer, conditioned on the active force being in direction $\ee_\chi$. These quantities obey the following equation, obtained by multiplying Eq.~\eqref{eq:master} by $\eta_{\XX_t+\rr}$ and summing over all tracer positions  and lattice configurations:
\begin{align}
&2d\tau^* {\partial_t k_{\rr}^{(\chi)}} = \sum_\mu(\nabla_\mu - \delta_{\rr,\ee_\mu}\nabla_{-\mu})k_{\rr}^{(\chi)} + \frac{\alpha}{2d-1}\sum_{\chi' \neq \chi} k_{\rr}^{(\chi')} \nonumber\\
&- \alpha k_{\rr}^{(\chi)}+\dfrac{2d\tau^*}{\tau}\sum_{\mu} p_\mu^{(\chi)}\moy{(1-\eta_{\XX_t+\ee_\mu})\nabla_\mu\eta_{\XX_t+\rr}}_\chi
\label{eq:evo_k}
\end{align}
 where $\nabla_\mu k_{\rr}^{(\chi)} = k_{\rr + \ee_\mu}^{(\chi)} - k_{\rr}^{(\chi)}  $ is a discrete gradient operator.
 
The evolution  equation  for the density profiles $k_{\rr}^{(\chi)}$ is not closed because it involves the correlation functions $\moy{\eta_{\XX_t+\ee_\mu}\eta_{\XX_t+\rr}}_\chi$. Obtaining an evolution equation for these correlation functions would actually involve correlation functions of higher order, and so on. We then resort to a decoupling approximation \cite{Illien2017c, Benichou2014, rizkallah2022microscopic},  which  consists in neglecting second order fluctuations around the mean, implying $\moy{\eta_{\XX_t+\ee_\mu}\eta_{\XX_t+\rr}}_\chi \simeq k_{\ee_\mu}^{(\chi)} k_{\rr}^{(\chi)}$. This enables to close Eq.~\eqref{eq:evo_k} and to obtain the following set of equations obeyed by the density profiles $k_{\rr}$:
\begin{align}
    2d\tau^*\partial_t k_{\rr}^{(\chi)} = & \sum_\mu A_\mu^{(\chi)} (\nabla_\mu + \delta_{\rr,\ee_\mu}) k_{\rr} \nonumber\\
    & - \alpha k_{\rr}^{(\chi)} + \frac{\alpha}{2d-1}\sum_{\chi' \neq \chi} k_{\rr}^{(\chi')}.
    \label{eq:eqk_decoupling}
\end{align}
We denoted $A_\mu^{(\chi)} \equiv 1 + \frac{2d\tau^*}{\tau} p_\mu^{(\chi)} [1-k_{\ee_\mu}^{(\chi)}]$ and adopted the convention $k_{\zz}^{(\chi)} = 0$. Note that this decoupling approximation preserves the spatial dependencies of the density profiles, and goes beyond trivial mean field, which would consist in writing $\moy{\eta_{\rr}} = \rho$ for any $\rr$.

We provide in SM \cite{SM} the stationary solution of Eq. \eqref{eq:eqk_decoupling}, which allows us to write explicitly the density profiles $k_{\rr}$  in terms of their values at the sites in the vicinity of the tracer $k_{\ee_\nu}$. In turn, these values are shown to satisfy a closed system of equations.  This finally provides the analytical solution of the stationary profiles  (up to the numerical solution of this implicit system of  equations) and of the stationary velocity  [Eq. \eqref{eq:speed}].
For a given value of the density, when the active force $F_A$ and the characteristic jump time of the bath particles $\tau^*$ are small enough, the average velocity of the tracer remains positive at all values of the external force $F_E$. Howewer, we observe that, for a sufficiently large persistence time $\tau_\alpha$, when the active force is large enough or when the bath particles are sufficiently slow compared to the tracer ($\tau^*\gg \tau$), the velocity can become a negative function of the external force (Fig. \ref{fig:decMC}(a)), which is the signature of ANM. We compare the value of the velocity predicted by our analytical theory with results from  Monte Carlo simulations of the microscopic dynamics, and observe an excellent agreement, which confirms the relevance of our decoupling approximation to  study the  dynamics of an active, driven tracer.

Finally, our analytical framework, that fully accounts for the microscopic details of the environment of the tracer, allows us to quantify the perturbation induced by its displacement. Indeed, as a byproduct of our calculation, we compute from Eq.~\eqref{eq:eqk_decoupling} the density profiles in the reference frame of the tracer. Their typical spatial dependence is plotted in Fig. \ref{fig:decMC}(b). It shows that ANM has a signature on the response of the environment, and that a small density excess may develop behind the tracer ($r<0$) when its average velocity becomes negative.

\begin{figure}
\begin{center}
\includegraphics[width = 0.7\columnwidth]{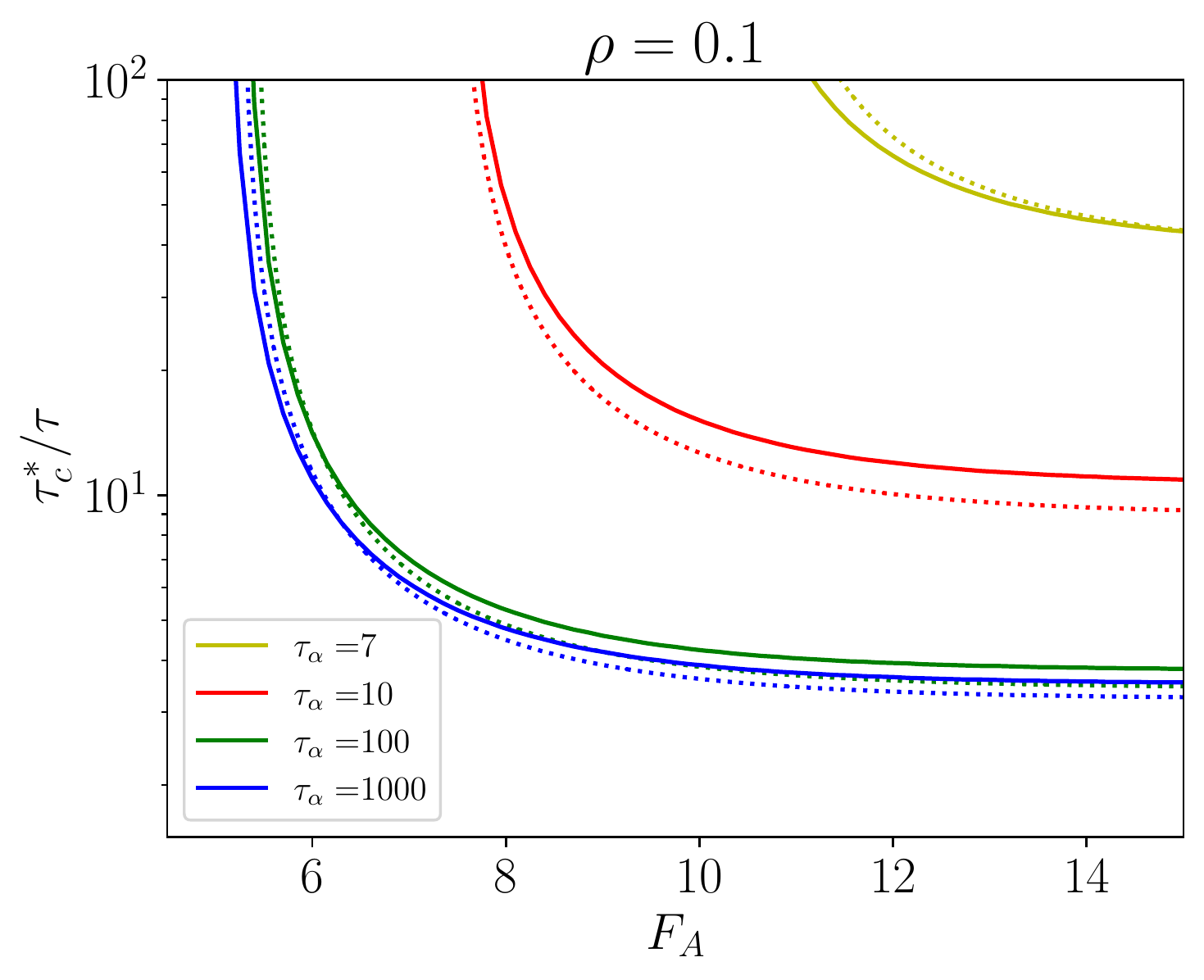}
\caption{Phase diagram for ANM. Above the lines, the theory predicts absolute negative mobility. Solid lines: decoupling approximation [Eqs. \eqref{eq:speed} and \eqref{eq:eqk_decoupling}]. Dotted lines: critical value $\tau^*_c/\tau$  determined from the low-density qualitative argument [Eq.~\eqref{eq:qualitativeSpeed}]. 
} 
\label{fig:phasediag}
\end{center}
\end{figure}

\emph{Phase diagram.---}
Relying on our analytical approach, we can explore a wide range of parameters to determine domains of existence of ANM. According to our previous studies on NDM \cite{Benichou2014,Benichou2015c}, for an infinitely persistent tracer (in the present formalism this corresponds to $F_A = 0$ and $F_E > 0$), this phenomenon occurs when, for a given value of the density, the ratio between the jump time of the obstacles and that of the tracer $\tau^*/\tau$ is sufficiently large. Here, the emergence of ANM is also determined by the parameters that control the activity of the tracer (the magnitude of the active force $F_A$ and the average persistence time of its orientation $\tau_\alpha$). In Fig. \ref{fig:phasediag}, we show the critical value of the characteristic jump time of bath particles $\tau^*_c$ (rescaled by $\tau$) above which ANM occurs, as a function of the active force $F_A$ for different values of the persistence time $\tau_\alpha$ and for a fixed value of the density of crowders $\rho$.

 \emph{Physical mechanism.---}  We now provide a physical interpretation of the phenomenon, which elucidates qualitatively the mechanism at the origin of ANM.
At low density of bath particles, the obstacles can be assumed to diffuse independently. For a given orientation of the active force $\chi$, the effective jump rate for the tracer is $\tau + \rho \tau_p^{(\chi)}$, where $\tau_p^{(\chi)}$ is the mean time that the tracer spends with a bath particles on one of its neighboring sites, and accounts for the `trapping' effect caused by the passive crowders. We can evaluate this typical time by considering that, when the tracer is at a given site $\RR$ and is blocked by a crowder located at site $\RR+\ee_1$, the tracer can move forward if one of these three independent events, which follow exponential laws, takes place: (i) the obstacle moves in a transverse direction with characteristic time $\frac{2d\tau^*}{(2d-2)}$; (ii) the active force changes direction with characteristic time $\tau_\alpha$; (iii) the tracer moves in a direction transverse to the direction of the obstacle with characteristic time $\tau/(1-p_1^{(\chi)}-p_{-1}^{(\chi)})$. The mean trapping time therefore follows an exponential law of characteristic time $\tau_p^{(\chi)}$ given by
\begin{equation}
    \frac{1}{\tau_p^{(\chi)}} = \frac{(2d-2)}{2d\tau^*} + \dfrac{1}{\tau_\alpha} + \frac{(1-p_1^{(\chi)} - p_{-1}^{(\chi)})}{\tau}.
    \label{eq:taup}
\end{equation}
The velocity of the tracer is then estimated as an average over the directions of active force $\chi$:
\begin{equation}
V\simeq \dfrac{1}{2d}\sum_\chi \dfrac{p_1^{(\chi)} -p_{-1} ^{(\chi)}}{\tau + \rho \tau_p^{(\chi)}},
\label{eq:qualitativeSpeed}
\end{equation}
and the condition for the existence of absolute negative mobility is  given by $\left. \frac{\dd V}{\dd F_E} \right|_{F_E=0} <0$.
Using the estimate of $\tau_p^{(\chi)}$ given by Eq.~\eqref{eq:taup}, we plot the critical value of the average jump time of the bath particles  $\tau_c^*$ above which ANM is expected as a function of the active force $F_A$ (Fig. \ref{fig:phasediag}), for different values of $\tau_\alpha$. This is compared to the result from the decoupling approximation, and it appears that our simple low-density argument is valid in a very wide range of parameters.

These physical considerations show how, in the low-density limit, the trapping of the tracer by passive crowders can result in ANM when its activity is strong enough. However, we emphasize that this approach would fail in dense regimes, where the correlations between the passive crowders would become predominant and would come into play. Moreover, the non-trivial response of the environment to the displacement of the tracer when ANM occurs (Fig. \ref{fig:decMC}(b)) cannot be predicted within this simplified framework.  The complete analytical solution presented above, which stems from the master equation, and which correctly captures these effects, is therefore necessary to fully describe the present problem.

\emph{Generalized Einstein relation.---} Finally, we derive from Eq. \eqref{eq:speed} the expression of the mobility of the tracer in the limit of small external force:
\begin{align}
&\lim_{F_E \to 0 }\frac{V}{F_E}  =   D_0 - \frac{1}{2 d \tau} \left[ (p_{1} - p_{-1})v_1  + \dfrac{2(2d - 1)\tau^*}{\alpha \tau} v_1^2 \right. \nonumber  \\
& \left. +2 \sum_{\epsilon \in \lbrace -1, 1 ,2 \rbrace} p_\epsilon\left(\gt_\epsilon -  \dfrac{\dd k_\epsilon}{\dd F_E}\right) (1 + (2d - 3)\delta_{\epsilon, 2})\right],
\label{eq:generalized_Einstein}
\end{align}
where $D_0 \equiv \lim_{t\to \infty}\frac{1}{2}\frac{\dd \moy{X_t^2}}{\dd t}$ is the diffusion coefficient of the active tracer without external force, that was previously determined through a higher-order decoupling approximation \cite{rizkallah2022microscopic}. In Eq. \eqref{eq:generalized_Einstein}, we used the shorthand notations $p_\epsilon = p_1^{(\epsilon)}$, $k_\epsilon = k_1^{(\epsilon)}$, and $v_1 = \sum_{\epsilon = \pm 1}   \epsilon p_\epsilon(1 - k_\epsilon)$, and we introduced the cross-correlations between the position of the tracer and the occupation of the sites in its vicinity $\gt_\epsilon = \langle (X_t - \moy{X_t}_\epsilon) \eta_{\XX_t + \ee_1}\rangle_\epsilon$.

The first term in Eq. \eqref{eq:generalized_Einstein} corresponds to the usual Einstein relation between the mobility of the tracer and its diffusion coefficient. The other term originates from the correlations between the displacement of the tracer and the dynamics of its environment. Importantly, its value controls the emergence of ANM: if it exceeds $D_0$ in absolute value, the mobility of the tracer becomes negative.
Therefore, our approach also provides an analytical expression for the generalized Einstein relation \cite{Baiesi2011,sarracino2016nonlinear}, that is derived explicitly from microscopic considerations.

\emph{Conclusion.---} We have shown that ANM can be observed in a minimal model for an active particle submitted to a constant external force  as a result of its interactions with the other particles in its environment. The analytical treatment of our microscopic theory provides an expression of the velocity of the tracer in this setting, allows us to determine the conditions for ANM to be observed, and gives insight into the response of the environment to the nonequilibrium dynamics of the tracer. 
Our framework can be applied to more complex geometries (channel-like systems, for instance), and could have interesting applications for selecting and sorting active tracers particles in crowded environments. 

\emph{Acknowledgments.---} A.S. warmly thanks Ralph Eichhorn for fruitful discussions, in particular on the qualitative argument for ANM. A.S. acknowledges partial support from MUR (Italian Ministry of University and Research) Project No.
PRIN201798CZLJ.


%

\end{document}


\title{Absolute negative mobility of an active tracer in a crowded environment \\ $\ $ \\ \textit{Supplementary Material}}

\author{Pierre Rizkallah}
\affiliation{Sorbonne Universit\'e, CNRS, Laboratoire de Physico-Chimie des \'Electrolytes et Nanosyst\`emes Interfaciaux (PHENIX), 4 Place Jussieu, 75005 Paris, France}

\author{Alessandro Sarracino}
\affiliation{Dipartimento di Ingegneria, Universit\`a della Campania Luigi Vanvitelli, 81031 Aversa (CE), Italy}

\author{Olivier B\'enichou}
\affiliation{Sorbonne Universit\'e, CNRS, Laboratoire de Physique Th\'eorique de la Mati\`ere Condens\'ee (LPTMC), 4 Place Jussieu, 75005 Paris, France}

\author{Pierre Illien}
\affiliation{Sorbonne Universit\'e, CNRS, Laboratoire de Physico-Chimie des \'Electrolytes et Nanosyst\`emes Interfaciaux (PHENIX), 4 Place Jussieu, 75005 Paris, France}

\date{\today}

\maketitle

\section{Numerical simulations}


We perform Monte Carlo numerical simulations on a 2D lattice with $M$ sites and initialize the $N$ bath particles in a random configuration,
with density $\rho = N/M$. We used a square lattice with $M = 200 \times 200$ sites, with periodic boundary conditions in both directions, and we checked that results are independent of the box size. Numerical results reported in the main text (Fig. 2) are obtained from averages over
about $10^3 - 10^4$ realizations.

\section{Analytical solution for density profiles}
We provide an analytical expression for the density profiles on an infinite $d-$dimensional lattice within our approximation scheme. We introduce the deviation to mean density $h_{\rr}^{(\chi)} \equiv k_{\rr}^{(\chi)} - \rho$ (defined in such a way that $\lim_{|\rr|\to\infty} h_{\rr}=0$) with the convention $h_{\zz}^{(\chi)} = 0$. The evolution equation is:
\begin{equation}
  2d\tau^*\partial_t h_{\rr}^{(\chi)} =  \sum_\mu A_\mu^{(\chi)}(\nabla_\mu + \delta_{\rr,\ee_\mu})   h_{\rr}^{(\chi)} + \sum_\mu\delta_{\rr,\ee_\mu} \rho(A_\mu-A_{-\mu}) 
- \alpha h_{\rr}^{(\chi)} + \frac{\alpha}{2d-1}\sum_{\chi' \neq \chi} h_{\rr}^{(\chi')}, \label{eq:h_Euler}  
\end{equation}
with $A_\mu^{(\chi)} \equiv 1 + \frac{2d\tau^*}{\tau} p_\mu^{(\chi)}[1-\rho-h_{\ee_\mu}^{(\chi)}]$.
We take the Fourier transforms, leading to an equation on the $2d$-dimensional vector $\boldsymbol{H}(\qq) \equiv \left( H^{(1)}(\qq), H^{(-1)}(\qq),\dots \right)$ where $H^{(\chi)}(\qq;t) \equiv \sum_{\rr} \ex{\ii\qq\cdot\rr} h_{\rr}^{(\chi)}(t)$: 
\begin{align}
2d\tau^* \partial_t H^{(\chi)}(\qq;t) = & K^{(\chi)}H^{(\chi)}(\qq;t)+K_0^{(\chi)}(\qq;t)-\alpha H^{(\chi)}(\qq;t)+\frac{\alpha}{2d-1}\sum_{\chi'\neq\chi}H^{(\chi')}(\qq;t). \label{eq:FourierH}
\end{align}
If we denote $h_\mu^{(\chi)} = h_{\ee_\mu}^{(\chi)}$, $q_\mu = \qq\cdot\ee_\mu$ and $\ee_{-\mu} = - \ee_\mu$, the functions involved in the Fourier transform are:
\begin{eqnarray}
K^{(\chi)}(\qq;t) & = & \sum_{\mu}     \left(\ex{-\ii q_\mu} -1\right)A_\mu^{(\chi)}(t),\\
K_0^{(\chi)}(\qq;t) & = &  \sum_{\mu}     \left(\ex{\ii q_\mu} -1\right)A_\mu^{(\chi)}(t) h_{\mu}^{(\chi)}(t) + 2\ii\sum_{j=1}^d \left[A_j^{(\chi)} -A_{-j}^{(\chi)} \right]\sin ( q_j).
\end{eqnarray} 
In the stationary regime, Eq. \eqref{eq:FourierH} can be written in the $2d \times 2d$ matrix form:
\begin{equation}
    \MM(\qq) \mathbf{H}(\qq)+ \mathbf{S}(\qq) = 0,
    \label{eq:matH}
\end{equation}
with the matrix $\MM(\qq)$ indexed by $(\chi, \chi') \in \lbrace \pm 1, ..., \pm d \rbrace^2$ and the $2d$ dimensional vector $\mathbf{S}$ defined by:
\begin{align}
    [\boldsymbol{M}(\qq)]_{\chi,\chi'} & = \left(-\alpha + \sum_{\mu}   \left(\ex{-\ii q_\mu} -1\right)A_\mu^{(\chi)})\right)\delta_{\chi,\chi'} + \dfrac{\alpha}{2d-1} (1-\delta_{\chi,\chi'}),\\
    \mathbf{S}(\qq)_\chi & = \sum_{\mu}     \left(\ex{\ii q_\mu} -1\right)A_\mu^{(\chi)}(t) h_{\mu}^{(\chi)}(t) + 2\ii\sum_{j=1}^d \left[A_j^{(\chi)} -A_{-j}^{(\chi)} \right]\sin ( q_j).
\end{align}
By inverting this matricial equation $\mathbf{H}(\qq) = -\MM^{-1}(\qq)\mathbf{S}(\qq)$, we get an expression of $H^{(\chi)}$ in function of the quantities $\left(h_\mu^{(\chi)}\right)_{(\mu, \chi) \in \lbrace \pm 1, ..., \pm d \rbrace^2}$ and the parameters of the problem. By taking the inverse Fourier transform of $H^{(\chi)}$ at the point $\ee_\mu$, we get the following equation: 
\begin{equation}
    h_\mu^{(\chi)} = - \int_{[-\pi,\pi]^d} \frac{\dd\qq }{(2\pi)^d}\ex{-i \qq \cdot \ee_\mu} [\MM^{-1}(\qq)\mathbf{S}(\qq)]_\chi
\end{equation}
This is a non linear equation (because $h_\mu^{(\chi)}$ also appears in $A_\mu^{(\chi)}$) involving only the densities in the vicinity of the tracer $\left(h_\mu^{(\chi)}\right)_{(\mu, \chi) \in \lbrace \pm 1, ..., \pm d \rbrace^2}$ and the parameters of the model. At first sight, it may seem that we need to perform $(2d)^2$ Fourier integrals in order to get enough equations to solve for the $h_\mu^{(\chi)}$, but in fact the number of independent unknowns is reduced to $10$ for $d = 2$ and $11$ for $d \geq 3$. We have the following symmetries:
\begin{align}
    h_2^{(1)} & = h_{-2}^{(1)}, \\
    h_2^{(-1)} & = h_{-2}^{(-1)}, \\
    h_\chi^{(\chi)} & = h_{2}^{(2)} &&\text{ for any } \chi \neq \pm 1, \\
    h_{-\chi}^{(\chi)} & = h_{-2}^{(2)} &&\text{ for any } \chi \neq \pm 1, \\ 
    h_{\mu}^{(\chi)} & = h_{\mu}^{(2)} &&\text{ for any } \chi \neq \pm 1, \mu = \pm 1, \\ 
    h_{\mu}^{(\chi)} & = h_{3}^{(2)} &&\text{ if } d \geq 3 \text{ for any } \chi \neq \pm 1, \mu \neq \pm 1, \pm \chi . 
\end{align}
Therefore, the following nonlinear system is closed and constitutes the analytical solution of the equations for the mean density in the vicinity of the tracer in the permanent regime, within our approximation scheme: 
\begin{equation}
    \boxed{h_\mu^{(\chi)} = - \int_{[-\pi,\pi]^d} \frac{\dd\qq }{(2\pi)^d}\ex{-i \qq \cdot \ee_\mu} [\MM^{-1}(\qq)\mathbf{S}(\qq)]_\chi \text{, for } (\chi, \mu) \in \lbrace \pm 1 \rbrace \times \lbrace \pm 1, 2 \rbrace \cup \lbrace 2 \rbrace \times \lbrace \pm 1, \pm 2, 3 \rbrace}.
    \label{eq:solution}
\end{equation}
Finally, to get the whole density profiles once the system is solved and the $h_\mu^{(\chi)}$ are known, we use $\mathbf{H}(\qq) = -\MM^{-1}(\qq)\mathbf{S}(\qq)$ (the right-hand side only depends on $\left(h_\mu^{(\chi)}\right)_{(\mu, \chi) \in \lbrace \pm 1, ..., \pm d \rbrace^2}$ and the parameters of the problem) from where we can compute any $h_{\rr}^{(\chi)}$ by taking the inverse Fourier transform:
\begin{equation}
    \boxed{h_{\rr}^{(\chi)} = - \int_{[-\pi,\pi]^d} \frac{\dd\qq}{(2\pi)^d} \ex{-i \qq \cdot {\rr}}[\MM^{-1}(\qq)\mathbf{S}(\qq)]_\chi}.
\end{equation}